\documentclass[preprint,preprintnumbers,showkeys,amsmath,amssymb,pra]{revtex4}

\usepackage{epsf}
\usepackage{graphicx}
\usepackage{dcolumn}
\usepackage{bm}
\usepackage{color}
\everymath{\displaystyle}
\begin{document}

\title{Field of a moving locked charge in classical electrodynamics}

\author{Alexander J. Silenko}
\email{alsilenko@mail.ru} \affiliation{Institute of Modern Physics, Chinese Academy of
Sciences, Lanzhou 730000, China,\\Bogoliubov Laboratory of Theoretical Physics, Joint Institute for Nuclear Research,
Dubna 141980, Russia,\\Research Institute for
Nuclear Problems, Belarusian State University, Minsk 220030, Belarus}

\date{\today}
\begin {abstract}
The paradox of a field of a moving locked charge (confined in a closed space) is considered and solved with the use of the integral Maxwell equations. While known formulas obtained for \emph{instantaneous} fields of charges moving along straight and curved lines are fully correct, measurable quantities are \emph{average} electric and magnetic fields of locked charges. It is shown that the \emph{average} electric field of locked charges does not depend on their motion. The average electric field of protons moving in nuclei coincides with that of protons being at rest and having the same spatial distribution of the charge density. The electric field of a twisted electron is equivalent to the field of a centroid with immobile charges which spatial distribution is defined by the wave function of the twisted electron.
\end{abstract}

\keywords{classical electromagnetism; beams in accelerators; twisted electron}

\maketitle


It has been shown in Ref. \cite{JPhysG2015locked} that the derivation of an electric field of a moving locked charge may lead to paradoxes. Such charges are confined in closed spaces like electrons in atoms, protons in nuclei, and charged particles in storage rings. It follows from the Lorentz transformations that the electric (and magnetic) field created by a charge significantly depend on its motion. In particular, the Lorentz transformation of the scalar potential has the form
\begin{eqnarray}\Phi=\frac{\Phi_0}{\sqrt{1-\bm\beta^2}}=\gamma\Phi_0,\qquad\bm\beta=\frac{\bm v}{c},\label{eqPhi}\end{eqnarray} where
$\Phi_0$ is the scalar potential in the rest frame and $\bm v$ and $\gamma$ are the velocity and Lorentz factor in the laboratory frame. When the particle moves along the $x$ axis, the corresponding Lorentz transformation of the electric field is given by \cite{LL2,Thirring,VanBladel}
\begin{eqnarray}
E_x=E_{0x},\qquad E_y=\frac{E_{0y}}{\sqrt{1-\bm\beta^2}}=\gamma E_{0y},\qquad E_z=\frac{E_{0z}}{\sqrt{1-\bm\beta^2}}=\gamma E_{0z},
\label{eqelfil}\end{eqnarray}
where $\bm E_0$ and $\bm E$ denote the electric field in the rest and laboratory frames, respectively. Certainly, electrodynamics presents general formulas for \emph{instantaneous} electric and magnetic fields defined by the Li\'{e}nard--Wiechert potentials \cite{LL2,VanBladel}. The \emph{instantaneous} fields of a charge moving along a circle have been given in the books \cite{Thirring,VanBladel} (see also Ref. \cite{Ruhlandt}).

The differences between $\Phi$ and $\Phi_0$ and $\bm E$ and $\bm E_0$ condition a difference between electric fields in the laboratory and rest frames [see Eqs. (\ref{eqfil}) and (\ref{eqfilwn}) below] and creates an illusion of a distinction between electric fields of a moving \emph{locked} charge and the corresponding charge at rest. In particular, this situation takes place for protons in nuclei and  particles and nuclei in storage rings (specifically, for beam--beam interactions) \cite{JPhysG2015locked}. It seems that the electric field should be bigger in absolute value than the field of the corresponding point charge. Naive averaging $\Phi$ and $\bm E$ over a charge trajectory leads to the relations $\overline{\Phi}\neq\Phi_0$ and $\overline{\bm E}\neq\bm E_0$ which strongly contradict to atomic and nuclear physics. This consideration clearly shows the presence of the paradox. 

We should mention that the analyzed paradox differs from the Schiff paradox \cite{VanBladel,Schiffparadox} and some other paradoxes explained in Ref. \cite{VanBladel}. In particular, the Schiff paradox is caused by a difference between the Maxwell equations in the rotating frame and the Minkowski space.

Beam--beam interactions
are important for electric-dipole-moment experiments in storage rings. In these experiments,
one can simultaneously use two beams consisting of particles moving in the clockwise and counterclockwise
directions with the same momentum \cite{dEDM,RevSciInstrum}. When the two beams are separated in space \cite{dEDM},
particles of one beam create a vertical electric field and a radial magnetic one acting on
particles of another beam. However, it has been shown in Ref. \cite{JPhysG2015locked} that their electric fields are defined by the Coulomb law even in the case of high velocities of nucleons and particles. In the present study, we give a more clear explanation of this paradox.

This explanation uses the two integral Maxwell equations:
\begin{eqnarray}
\oint{\bm E\cdot d\bm S}=4\pi\int{\varrho dV},
\label{eqinton}\end{eqnarray}
\begin{eqnarray}
\oint{\bm B\cdot d\bm l}=\frac{4\pi}{c}\int{\left(\bm j+\frac{1}{4\pi}\frac{\partial \bm E}{\partial t}\right)\cdot d\bm S}.
\label{eqinttw}\end{eqnarray}
In the first equation, $S$ is a closed surface limiting the volume $V$ and $\varrho$ and $\bm j$ are the charge and current densities. In the second equation, $l$ is a closed line limiting the surface $S$.

The key point of the present explanation is an independence of the \emph{average} charge density of charged particles, nucleons, and nuclei moving along a definite trajectory from the velocity of their motion, $\bm v$. This independence is conditioned by the charge conservation law. Evidently, the total charge does not depend on the velocity. Therefore, $\overline{\varrho}=\varrho_0$ and $\overline{\bm j}=\overline{\varrho}\bm v=\varrho_0\bm v$ \emph{in any point} of the particle (nucleon, nucleus) trajectory. Here $\varrho_0$ is the average charge density of slowly (nonrelativistically) moving particles and nuclei.
The use of Eqs. (\ref{eqinton}) and (\ref{eqinttw}) simplifies an analysis of distinctive features of the fields of locked charges. Let us first compare fields created by beams of charged particles or nuclei in a storage ring and in a free space. We can suppose the beam radii to be rather small as compared with the distance to the beams. In this case, the beams can be considered as finite or infinite thin filaments. The electric field of an infinite filament at rest is defined by
\begin{eqnarray}
\bm E=E\frac{\bm r}{r}, \qquad E=\frac{2\tau}{r},
\label{eqfil}\end{eqnarray} where $r$ is the distance to the filament and $\tau$ is the charge of the unit of its length. When the filament moves, its electric field is given by
\begin{eqnarray}
E=\frac{2\tau\gamma}{r}, \label{eqfilwn}\end{eqnarray}
where
$\gamma$ is the Lorentz factor.
This significant difference has been noted in Ref. \cite{JPhysG2015locked}.

The motion of charges
creates also an electric current which magnetic field is given by
\begin{eqnarray}
\bm B=\frac{2\tau\gamma\bm\beta\times\bm r}{r^2}.
\label{eqfilwm}\end{eqnarray}

When a filament is finite but its length satisfies the condition $l\gg r$ and observation points are far from the ends of the filament, the electric and magnetic fields are also defined by Eqs. (\ref{eqfilwn})
and (\ref{eqfilwm}). Certainly, Eqs. (\ref{eqinton}) and (\ref{eqinttw}) lead to the same results because of the Lorentz contraction of the filament length. Therefore, the fields nearby the moving filament are proportional to the Lorentz factor $\gamma$ under the above-mentioned conditions. This situation takes place for charged filaments in both the free space and a storage ring (if a curvature of a beam trajectory can be neglected). We should underline that all results previously obtained for \emph{instantaneous} fields of charges moving along straight and curved lines are fully correct. However, only \emph{average} electric and magnetic fields of locked charges are measurable.

It has been ascertained in Ref. \cite{JPhysG2015locked} that the average electric and magnetic fields created by a charged beam in a storage ring are defined by the following equations:
\begin{eqnarray}
\overline{\bm E}=\frac{2\overline{\tau}\bm r}{r^2},
\label{eqfilng}\end{eqnarray}
\begin{eqnarray}
\overline{\bm B}=\frac{2\overline{\tau}\bm\beta\times\bm r}{r^2}. \label{eqfilwnng}\end{eqnarray}

This paradox has a clear explanation. The Lorentz contraction of the length of the moving filament increases the \emph{local} charge density $\gamma$ times. This effect takes also place for particle bunches in a storage ring, and their \emph{instantaneous} fields satisfy Eqs. (\ref{eqfilwn}) and (\ref{eqfilwm}). Nevertheless, the \emph{average} charge density of charges confined in a closed space is constant and cannot be changed by their motion. As a result, the \emph{average} fields of moving charges are defined by Eqs. (\ref{eqfilng}) and (\ref{eqfilwnng}) for average quantities. The latter equation follows from the fact that the average \emph{linear} densities of the charge and current ($\overline{\tau}$ and $\overline{\bm i}$, respectively) satisfy the relation
\begin{eqnarray}
\overline{\bm i}=\overline{\tau}\bm v. \label{jvrho}\end{eqnarray}
Certainly, the average value of the quantity $\partial \bm E/(\partial t)$ is zero.

To expound this situation and to check a dependence of the average electric and magnetic fields on the velocity of the finite filament, let us specify that the length of the filament at rest is $l_0$. While the field of the relativistically moving filament is equal to $E=\gamma E_0$, its length is correspondingly shorter ($l=l_0/\gamma$). The average field of the moving filament is defined by its linear charge density when the filament charge, $Q=\tau l$, is extended on the beam circumference $C$. Therefore, $\overline{\tau}=Q/C=\tau l/C=\overline{\tau_0}$ and $\overline{\bm i}=\overline{\tau}\bm v=\overline{\tau_0}\bm v$.

In quantum mechanics, one can consider stationary states of a charged particle in the storage ring. For such states, the charge density is proportional to the square of a wave eigenfunction and is almost independent of the azimuth $\phi$ defined relative to the center of the storage ring. In any stationary state, the particle is extended on the whole beam path and a passage to the average charge density is well substantiated. Average values (i.e., expectation values) of the particle momentum in the Cartesian coordinates are equal to zero in any stationary state. Particles and nuclei in storage rings are usually (but not always) uniformly distributed along the beam trajectory, e.g., due to the momentum spread. In this case, the charge density is stationary, uniform, and independent of the beam velocity. The fields need not be averaged and satisfy Eqs. (\ref{eqfilng}) and (\ref{eqfilwnng}). Even if the charged particle is single, one can consider its stationary states in the storage ring. In any stationary state, the fields are stationary and are defined by Eqs. (\ref{eqfilng}) and (\ref{eqfilwnng}). In the classical limit, one passes to classical wave theory and describes particles and nuclei by de Broglie waves. The use of such a description in the stationary state can also simplify the explanation of the paradox of the electric and magnetic fields created by a moving particle or nucleus in the storage ring. The corresponding de Broglie wave fills the whole beam path and its momentum is equal to zero ($\bm P=\overline{\bm p}=0$), while $P_\phi\neq0$. We can use the integral Maxwell equations (\ref{eqinton}) and (\ref{eqinttw}) and can check the independence of the electric field strength of the particle/nucleus velocity for any closed surface $S$ containing the storage ring.

We can conclude that passing to the charge and current densities sheds light on the origin of the paradox investigated. When the charged filament or the single charge moves, its electric field increases as compared with the static case due to the length contraction. This effect also takes place for the closed filament and charge but only for periods of time when these sources are close to the observation point. Averaged electric fields of moving sources do not depend on their velocities because their averaged charge densities remain the same.

We can now consider the electric field of a proton moving in a nucleus. Certainly, the moving proton undergoes the Lorentz contraction and its charge density increases $\gamma$ times. As a result, the scalar potential and the electric field strength of its field increase accordingly Eqs. (\ref{eqPhi}) and (\ref{eqfilwn}). However, the \emph{average} charge density of the proton, $\overline{\varrho}$, does not depend on its motion in a fixed closed space. Indeed, the alternative statement $\overline{\varrho}=\overline{\gamma}\,\overline{\varrho_0}$ is absurd because its integration leads to a noninvariance of the proton charge.

Therefore, the average electric field of protons moving in nuclei, in agreement with Ref. \cite{JPhysG2015locked}, coincides with that of protons being at rest and having the same spatial distribution of the charge density. In particular, the electric field of a spherically symmetric nucleus is exactly defined by the Coulomb law in agreement with the real situation.

The results obtained bring a rather important conclusion. The nucleus mass is affected by a motion of protons and neutrons but the electric field of the nucleus is independent of their motion and can be obtained by a summation of electric fields of charges (protons) \emph{at rest}. The electric and magnetic fields of a moving \emph{nucleus} are defined by the Lorentz transformation from its rest frame and coincide with fields of a nucleus with the same charge density but with charges at rest.

The same conclusions are valid for atoms and other closed systems. Amazingly, they are also applicable for charged twisted (vortex) particles while they can be modeled by wave beams or wave packets moving with certain velocities. Twisted particles possess an intrinsic orbital angular momentum (OAM). Twisted electron beams with large intrinsic OAMs (up to 1000$\hbar)$ have been recently obtained \cite{VGRILLO}. Since twisted electrons possess large magnetic moments [see Eq. (\ref{mdm}) below], their discovery has
opened new possibilities in the electron microscopy and investigations of magnetic phenomena
(see Refs. \cite{BliokhSOI,Lloyd,Rusz,Edstrom,imaging,Observation,OriginDemonstration,LarocqueTwEl}
and references therein). The twisted electron is not a pointlike particle. Its spatial distribution is localized in the transversal plane. Such an electron can be modeled by a charged centroid being an extended object with a charge distribution defined by the wave function of the twisted electron \cite{BliokhSOI,ResonanceTwistedElectrons}. The centroid as a
whole is characterized by a center-of-charge radius vector and by a kinetic momentum. An intrinsic motion of partial charges takes place. The twisted electron possesses a tensor magnetic polarizability \cite{PhysRevLettEQM2019} and a measurable (spectroscopic) electric quadrupole moment \cite{PhysRevLettEQM2019,KarlovetsZhevlakov} and its interactions with external electric and magnetic fields lead to nontrivial effects  (see, e.g., Refs. \cite{ResonanceTwistedElectrons,PhysRevLettEQM2019,Bliokhmagnetic,Manipulating,snakelike}). 
The effective (kinematic) mass of a twisted electron manifesting in experiments is bigger than the electron rest mass \cite{KarlovetsParaxial,LightArXiv}. The effective mass corresponds to the total electron energy in the rest frame of the charged centroid. This energy is equal to the sum of the rest energy and the kinetic energy of a hidden motion of the electron and is quantized \cite{LightArXiv}. The inertial and kinematic masses of the twisted electron always coincide \cite{LightArXiv}. Beams of free twisted electrons can be presented as ensembles of plane waves and, certainly, their motion is not accelerated.

The model of the charged centroid simplifies a consideration of an average electric field created by the twisted electron.
Our analysis shows that this field (like the electric field of atomic electrons) is equivalent to the field of \emph{immobile} charges which spatial distribution is defined by the wave function of the twisted electron. The total charge of these immobile charges is equal to $e$.

We can conclude that the analysis of the average electric field created by the twisted electron confirms its modeling by the charged centroid with the kinematic and inertial masses coinciding with each other and different from the electron rest mass. In the \emph{centroid rest frame}, this field is equivalent to the above-mentioned field of \emph{immobile} charges.

A motion of the centroid is not the only source of the magnetic field of the twisted electron. Any charge motion creates an electric current. Therefore, the intrinsic motion of partial charges forming the centroid conditions both the OAM $\bm L$ and the orbital magnetic moment $\bm\mu_L$. The universal connection between the orbital magnetic moment $\bm\mu_L$ and the OAM $\bm L$ of a free \emph{pointlike} particle has the form \cite{Barut}
$$\bm\mu_L=\frac{e\bm L}{2E}$$ in any frame. Here $E$ is the total energy of the twisted electron. The magnetic moment is also contributed by the spin. If the Dirac spin magnetic moment of the electron is taken into account and the small anomalous magnetic moment is disregarded, the total magnetic moment reads \cite{ResonanceTwistedElectrons}
\begin{equation}
\begin{array}{c}
\bm\mu=\bm\mu_L+\bm\mu_s=\frac{e(\bm L+2\bm s)}{2E}.
\end{array}
\label{mdm}
\end{equation}

In summary, we have considered the paradox of a field of a moving locked charge stated in Ref. \cite{JPhysG2015locked} and have definitively solved it. Known formulas \cite{LL2,Thirring,VanBladel} describing \emph{instantaneous} electric and magnetic fields of charges moving along straight and curved lines are fully correct. However, measurable quantities are \emph{average} electric and magnetic fields of locked charges. The average charge density of charged particles, nucleons, and nuclei moving along a definite trajectory does not depend on the velocity of their motion. With the use of the integral Maxwell equations, we have shown that the \emph{average} electric field of locked charges does not depend on their velocity either. This is the main result obtained in the present paper. We have analyzed the electric fields of protons moving in nuclei and of the twisted electron. The former field coincides with that of protons being at rest. The latter field is equivalent to the field of a centroid with immobile charges which spatial distribution is defined by the wave function of the twisted electron.

\section*{Acknowledgements}

This work was supported by the Belarusian Republican Foundation
for Fundamental Research
(Grant No. $\Phi$18D-002), by the National Natural Science
Foundation of China (Grant No. 11575254), and
by the National Key Research and Development Program of China
(No. 2016YFE0130800).
A. J. S. also acknowledges hospitality and support by the
Institute of Modern
Physics of the Chinese Academy of Sciences.


\end{document}